\documentclass[aip,jmp,reprint,amsmath,amssymb,onecolumn]{revtex4-1}
\usepackage{color}

\usepackage[utf8]{inputenc}
\usepackage{graphicx}
\usepackage{dcolumn}
\usepackage{bm}
\usepackage{subfigure}
\usepackage{amssymb}
\usepackage{tabularx}
\usepackage{amsmath}
\usepackage{ulem}
\usepackage{xspace}

\begin{document}

\title{Time Reversal Invariance of quantum kinetic equations II: Density operator formalism}

\author{Michael Bonitz, Miriam Scharnke, and Niclas Schlünzen}

 \affiliation{%
     Christian-Albrechts-University Kiel, Institute for Theoretical Physics and Astrophysics, Leibnizstra\ss{}e 15, 24098 Kiel, Germany
 }% 

\date{\today}

\begin{abstract}
  \noindent Time reversal symmetry is a fundamental property of many quantum mechanical systems. The relation between statistical physics and time reversal is subtle and not all statistical theories conserve this particular symmetry, most notably hydrodynamic equations and kinetic equations such as the Boltzmann equation. Here we consider quantum kinetic generalizations of the Boltzmann equation by using the method of reduced density operators leading to the quantum generalization of the BBGKY-(Bogolyubov, Born, Green, Kirkwood, Yvon) hierachy. We demonstrate that all commonly used approximations, including Vlasov, Hartree-Fock and the non-Markovian generalizations of the Landau, T-matrix and Lenard-Balescu equations are originally time-reversal invariant, and we formulate a general criterion for time reversibility of approximations to the quantum BBGKY-hierarchy. Finally, we illustrate, on the example of the Born approximation, how irreversibility is introduced into quantum kinetic theory via the Markov limit, making the connection with the standard Boltzmann equation. This paper is a complement to paper I [Scharnke {\it et al.}, submitted to J. Math. Phys., arXiv:1612.08033] where time-reversal invariance of quantum-kinetic equations was analyzed in the frame of the independent nonequilibrium Green functions formalism. 
\end{abstract}

\maketitle
\date{\today}

\renewcommand{\i}{\mathrm{i}}
\renewcommand{\d}{\mathrm{d}}
\renewcommand{\H}{\mathrm{H}}
\renewcommand{\S}{\mathrm{S}}

\section{Introduction}
\noindent
%{\bf section not finished yet!}\\

The time evolution of quantum many-body systems is of high current interest in many areas of modern physics and chemistry for example in the context of laser-mater interaction, non-stationary transport or dynamics following an interaction or confinement quench. The theoretical concepts to study these dynamics are fairly broad and include (but are not limited to) wave function based approaches, density functional theory and quantum kinetic theory. The latter treats the time dynamics of the Wigner distribution or, more generally, the density matrix and captures the relaxation towards an equilibrium state (see, e.g. Refs.~\onlinecite{klimontovich_book, bogolyubov, spatschek, bonitz_qkt}). The most famous example of a kinetic equation is the Boltzmann equation, together with is quantum generalization, but this equation is known to be not applicable to the short-time dynamics. For this reasons generalized quantum kinetic equations were derive that are non-Markovian in nature (e.g. Refs.~\onlinecite{klimontovich_book, bonitz_pla96, bonitz_qkt, akbari,schuck_16, lacroix_prb_14, hermanns_prb_14}), and that have a number of remarkable properties including the conservation of total energy, in contrast to kinetic energy conservation in the Boltzmann equation. It was recently demonstrated that these generalized quantum kinetic equations are well suited to study the relaxation dynamics of weakly and moderately correlated quantum systems, in very good agreement with experiments with ultracold atoms (e.g. Refs.~\onlinecite{schluenzen16,schluenzen2}), and first-principle density matrix renormalization group methods~\cite{schluenzen17}.

This success of generalized quantum kinetic equations warrants a more detailed theoretical analysis of their properties. Despite extensive work over the recent decades the aspect of time reversibility was not studied in detail.
The relation between time reversal symmetry and statistical physics is generally subtle, and not all statistical theories are invariant under time reversal, the most famous counterexample being the above mentioned Boltzmann equation of classical statistical mechanics and its quantum generalization. 
%Therefore, extensive work has been done over the recent seven decades to derive 
In contrast, the non-Markovian generalizations of the Boltzmann equation which can be used to improve the Boltzmann equation and contain the latter as a limiting case are expected to be time-reversal invariant as the underlying quantum mechanical system. But then the question arises, where exactly time-reversal invariance is lost, how this is related to common many-body approximations and so on.

Among the well established approaches to derive these generalized quantum kinetic equations we mention density operator concepts, see e.g. Ref.~\onlinecite{bonitz_qkt} for an overview, and nonequilibrium Green functions (NEGF). The question of time-reversal invariance within the NEGF-formalism was recently analyzed by us in paper I \cite{scharnke_17}.
%\newline
%
It is the goal of the present article to complement the NEGF results of that paper by an analysis of the independent and technically very different density operator formalism. In this paper we briefly recall the derivation of the quantum BBGKY-hierarchy (Bogolyubov-Born-Green-Kirkwood-Yvon) in Sec.~\ref{s:bbgky}. Since the BBGKY-hierarchy can be directly derived from the Heisenberg equation (von Neumann equation) for the $N$-particle density operator which is time-reversal invariant, it should be expected that this hierarchy  has the same symmetry properties. 
Nevertheless, a general proof is usually missing in the literature, e.g. Refs.~\onlinecite{klimontovich_book, bogolyubov, spatschek, bonitz_qkt}, and a successful procedure is presented in Sec.~\ref{s:tri-bbgky}. We then demonstrate in Sec.~\ref{s:tri-approx} that important standard closure approximations to the BBGKY-hierarchy also preserve time reversal symmetry. 
In Sec.~\ref{s:born} we demonstrate, for an example, the transition from a time-reversal invariant generalized kinetic equation to an irreversible equation of the Boltzmann type, by performing the Markov limit and the weakening of initial conditions.
We conclude with a summary in Sec.~\ref{s:summary}.

%-----------------------------
\section{BBGKY-Hierarchy for the Reduced Density Operators}
\label{s:bbgky}
\noindent Here we briefly recall the basic equations of density operator theory following Ref.~\onlinecite{bonitz_qkt}.
The generic hamiltonian of an interacting $N$-particle system is given by a sum of a single-particle and an interaction term%
\begin{align}\label{hn_def}
%{\hat H}_{1\dots N} &=\sum\limits_{i=1}^{N}\;{\hat H}_i
{\hat H} &=\sum\limits_{i=1}^{N}\;{\hat H}_i
+\sum\limits_{1\le i < j \le N}\, {\hat V}_{ij},
\\
{\hat H}_i(t) &=\frac{{\hat p}^2_i}{2\,m_i} + {\cal {\hat U}}_i(t).
\label{hi-def}
\end{align}
The solutions of the time-dependent $N$-particle Schrödinger equation with this hamiltonian are denoted by $|\psi^{(1)}\rangle$ $\dots |\psi^{(M)}\rangle$ and form a complete orthonormal basis,
\begin{align}
\langle \psi^{(k)}|\psi^{(l)}\rangle &= \delta_{k,l},
\\
\sum\limits_{k=1}^{M}
|\psi^{(k)}\rangle \langle \psi^{(k)}| &= 1.
\label{on-basis}
\end{align}
The central quantity for the construction of quantum kinetic equations is the $N$-particle density operator, 
\begin{align}
 {\hat \rho}=\sum\limits_{k=1}^{M}\,W_k \,
|\psi^{(k)}\rangle \langle \psi^{(k)}|,
\label{eq:def-rho}
\end{align}
where the $W_k$ are positive real probabilities, $0\le W_k \le 1$, with $\sum_{k=1}^M W_k =1$, and we restrict ourselves to the case of time-independent probabilities. The density operator obeys the von Neumann equation
\begin{align}
 \i\hbar\frac{\partial}{\partial t} {\hat \rho} -
%[{\hat H}_{1\dots N},{\hat \rho}_{1\dots N}] = 0.
[{\hat H},{\hat \rho}] = 0.
\label{eq:von_neumann}
\end{align}
In order to derive the quantum BBGKY-hierarchy, we introduce the reduced $s$-particle density operator (${s=1\dots N-1}$)
\begin{align}
 {\hat F}_{1\dots s}= C_s^N \, \mbox{Tr}_{s+1 \dots N}\,{\hat \rho},
\quad
\mbox{Tr}_{1\dots s}{\hat F}_{1\dots s}=C_s^N, 
\label{eq:fs-def}
\end{align}
where $C_s^N = \frac{N!}{(N-s)!}$.
The equations of motion for the reduced density operators follow directly from the von Neumann equation (\ref{eq:von_neumann}) and the definition (\ref{eq:fs-def}),
\begin{equation}\label{eq:bbgky_finite}
%\fbox{$\D{
\i\hbar\frac{\partial}{\partial t} {\hat F}_{1\dots s}-[{\hat H}_{1\dots s},{\hat F}_{1\dots s}]=
 \mbox{Tr}_{s+1}\sum_{i=1}^{s}[{\hat V}_{i,s+1},{\hat F}_{1\dots s+1}],
%}$}
\end{equation}
where ${\hat H}_{1\dots s}$ is the $s$-particle Hamilton operator which
follows from the $N$-particle hamiltonian, Eq.~(\ref{hn_def}), by substituting $N\rightarrow s$. The system (\ref{eq:bbgky_finite}) with $s=1\dots N-1$  constitutes the quantum generalization of the BBGKY-hierarchy.	

In order to specify decoupling approximations to the hierarchy we introduce the correlation operators,
\begin{align}
%\fbox{$\D{
%\begin{array}{rcl}
{\hat F}_{12} &= {\hat F}_{1}{\hat F}_{2} + {\hat g}_{12},
\label{eq:g12-def}
\\
{\hat F}_{123} &= {\hat F}_{1}{\hat F}_{2}{\hat F}_{3} + {\hat g}_{23}{\hat F}_{1} + {\hat g}_{13}{\hat F}_{2}+ {\hat g}_{12}{\hat F}_{3} + {\hat g}_{123},
\label{eq:g123-def}
\end{align}
%}$}
%
where ${\hat g}_{12}$ describes pair correlations, ${\hat g}_{123}$ three-particle correlations and so on that are due to interaction effects beyond mean fied.
In contrast, mean field (Vlasov, Hartree-Fock) terms are contained in the products of single-particle density operators and appear via the mean field potential ${\hat U}^{\rm H}_i = \mbox{Tr}_j {\hat V}_{ij} {\hat F}_j$ leading to renormalization of the single-particle and two-particle hamiltonians 
${\hat H}_i \to {\hat {\bar H}}_i = {\hat H}_i + {\hat U}^{\rm H}_i$, 
${\hat H}_{ij} \to {\hat {\bar H}}_{ij} = {\hat {\bar H}}_i + {\hat {\bar H}}_j + {\hat V}_{ij}$ and so on. 
The BBGKY-hierarchy rewritten in terms of the correlation operators then becomes
\begin{align}
 \i \hbar\frac{\partial}{\partial t} {\hat F}_{1} - [{\hat {\bar H}_1},{\hat F}_1]
&=
\mbox{Tr}_{2}[{\hat V}_{12},{\hat g}_{12}],
\label{eq:f1-eq}
\\
\i\hbar \frac{\partial}{\partial t}
{\hat g}_{12} - [{\hat {\bar H}}_{12}, {\hat g}_{12}]
&= [{\hat V}_{12},{\hat F}_{1}{\hat F}_{2}]  \, +
\label{eq:g12-eq}
\\
+ \mbox{Tr}_{3}\Big\{ [{\hat V}_{13},{\hat F}_1 {\hat g}_{23}] &+ [{\hat V}_{23},{\hat F}_2 {\hat g}_{13}] +
[{\hat V}_{13}+{\hat V}_{23},{\hat g}_{123}]\Big\},
\nonumber
\end{align}
and similarly for the higher order operators.
Standard many-body approximations are easily identified from equations (\ref{eq:f1-eq}) and (\ref{eq:g12-eq}), cf. for example Ref.~\onlinecite{bonitz_qkt}:
\begin{enumerate}
 \item The {\it mean field (Hartree or Hartree-Fock) approximation} that leads to the nonlinear Vlasov equation (or to time-dependent Hartree-Fock) follows from letting ${\hat g}_{12} \to 0$, in Eq.~(\ref{eq:f1-eq}).
 \item The {\it second order Born approximation} leading to the Landau equation follows  from neglecting ${\hat V}_{12}$ in ${\hat {\bar H}_{12}}$ on the left 
and ${\hat g}_{23} = {\hat g}_{13} = {\hat g}_{123} \to 0$, on the right side in Eq.~(\ref{eq:g12-eq}).
  \item The {\it T-matrix or ladder approximation} follows from setting ${\hat g}_{23} = {\hat g}_{13} = {\hat g}_{123} \to 0$, on the right side in Eq.~(\ref{eq:g12-eq}).
  \item The {\it polarization approximation} that is related to the GW approximation of Green functions theory and leads to the Lenard-Balescu equation follows from neglecting ${\hat V}_{12}$ in ${\hat {\bar H}_{12}}$ on the left and ${\hat g}_{123} \to 0$, on the right side in Eq.~(\ref{eq:g12-eq}).
  \item The {\it screened ladder approximation} that is related to the parquet approximation (or ``FLEX'') in Green functions theory follows from ${\hat g}_{123} \to 0$, on the right side in Eq.~(\ref{eq:g12-eq}).  
\end{enumerate}
In similar manner, higher order decoupling schemes for the BBGKY-hierarchy are introduced on the level of the equation of motion for $g_{123}$. Typically, approximations are derived by omitting terms of the form $[{\hat A}, {\hat B}]$, where ${\hat A}$ is a contribution to the full hamiltonian (\ref{hn_def}) (typically an interaction potential) and ${\hat B}$ are contributions to the cluster expansion (\ref{eq:g123-def}). This will be discussed in more detail in Sec.~\ref{s:tri-approx}.

Finally, we note that the cluster expansion  (\ref{eq:g123-def}) is written without explicit account of the spin statistics. A direct (anti-)symmetrization of the hierarchy, for the case of bosons (fermions), is straightforwardly achieved by replacing the density operators according to \cite{boercker-etal.79}
\begin{align}
 {\hat F}_{1\dots s} \longrightarrow {\hat F}_{1\dots s} \Lambda^\pm_{1\dots s},
\end{align}
where the (anti-)symmetrization operators are given by
\begin{align}
\Lambda^\pm_{12} & = 1 \pm P_{12},
\nonumber\\  
\Lambda^\pm_{123} &= 1 \pm P_{12} \pm P_{13}  \pm P_{23} +  P_{12} P_{13} +  P_{12} P_{23},
\nonumber
\end{align}
and so on, where $P_{ij}$ is the permutation operator of particles $i$ and $j$ and the upper (lower) sign referes to bosons (fermions). (Anti-)symmetrization is then achieved by applying the $s$-particle operator $ \Lambda^\pm_{1\dots s}$ to the s-th equation of the BBGKY-hierarchy, term by term.
We illustrate this procedure for the (anti-)symmetrization of the Hartree mean field term, on the l.h.s. of Eq.~(\ref{eq:f1-eq}), which is obtained by replacing $\hat F_1 \hat F_2 \to \hat F_1 \hat F_2 \Lambda^\pm_{12}$,
\begin{align}
 [{\hat U}^{\rm H}_1, \hat F_1] & \longrightarrow [{\hat U}^{\rm HF}_1, \hat F_1] = \mbox{Tr}_2 [\hat V_{12}, \hat F_1 \hat F_2 \Lambda^\pm_{12}],
\nonumber\\
\mbox{with} \quad {\hat U}^{\rm HF}_i &= \mbox{Tr}_j \hat V_{ij} \hat F_j \Lambda^\pm_{ij},
\end{align}
The full (anti-)symmetrized equations are given in Ref.~\onlinecite{bonitz_qkt}. However, we will not need these equations below. The reason is that the (anti-)symmetrization operators commute with the time reversal operator ${\hat T}$, cf. Sec.~\ref{s:tri}. Therefore, (anti-)symmetrization does not affect the time reversal properties of the resulting equations and approximations, allowing us to restrict ourselves to the simpler equations (\ref{eq:f1-eq}) and (\ref{eq:g12-eq}), in the following.

%----------------------
\section{Time Reversal Invariance in Quantum Many-Body Theory}\label{s:tri}
\subsection{Time Reversal Invariance of the Equations of Motion of Quantum Mechanics}\label{sec:tri_se}
Let us recall the concept of time reversibility as was discussed in Ref.~\onlinecite{scharnke_17}, for text book discussions, see Refs.~\onlinecite{haake, zubarev}.
 Consider the time-dependent $N$-particle Schr\"odinger equation on an arbitrary finite interval of time, 
$-t_0 \le t \le 0$, with a given initial condition $|\psi_0\rangle$,
\begin{align}
 \label{eq:tdTSE}
 \i\hbar\,\partial_t|\psi(t)\rangle &= \hat{H}|\psi(t)\rangle\;,
\\
 |\psi(-t_0)\rangle &= |\psi_0\rangle.
\end{align}
This equation is called time reversal invariant if, 
\begin{description}
 \item[i] for any solution $|\psi(t)\rangle$, there exists another solution $|\psi'(t')\rangle$ with $t' \in [0, t_0]$ and $t' = -t$, and if
 \item[ii] there exists a unique relation between the two: 
   \begin{equation}
     |\psi'(t')\rangle = \hat{T}|\psi(t)\rangle,
   \label{eq:psi-psiprime}   
    \end{equation}
\end{description}
\begin{figure}
 \includegraphics[width=.5\linewidth]{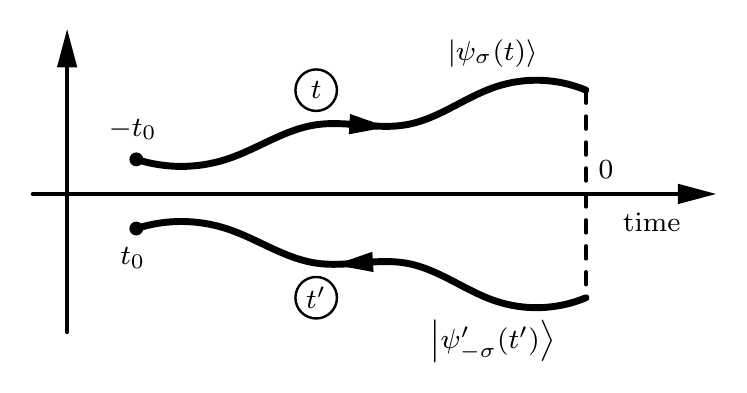}
 \caption{Illustration of the forward and backward solutions of the time-dependent Schrödinger equation. Upper trajectory: forward solution $|\psi_{\sigma}(t)\rangle$. Lower trajectory: backward solution $|\psi'_{-\sigma}(t')\rangle$. Note that we choose the limits of the forward trajectory as $t=-t_0$ and $t=0$, whereas the backward one runs from  $t`=0$ to $t`=t_0$. The time reversal occurs at $t=0$.
 }
 \label{fig:tri-tdse}
\end{figure}
where the time-reversal operator ${\hat T}$ will be specified below. Both solutions describe the same physical state, therefore, the associated probability densities must coincide, 
\begin{equation}
| |\psi_{\sigma}(t)\rangle |^2 = | |\psi_{-\sigma}'(-t)\rangle |^2,  
\label{eq:probability_reversal}
\end{equation}
where we indicated explicitly that, on the backward trajectory $|\psi'(t')\rangle$, the spin projections $\sigma$ of all particles are inverted. Analogously, momenta and angular momenta (their eigenvalues) are inverted, as in classical mechanics. To motivate the choice of ${\hat T}$, we rewrite the Schrödinger dynamics (\ref{eq:tdTSE}) in terms of the standard time-evolution operator ${\hat U}$,
\begin{align}
 |\psi(t)\rangle & = {\hat U}(t, -t_0)|\psi_0\rangle,
\\
{\hat U}(t, t') &= {\cal T} \, e^{ -\frac{\i}{\hbar}\int\limits_{t'}^t \d{\bar t}\,{\hat {\bar H}}({\bar t})  }.
\label{eq:un-solution}
\end{align}
Backward evolution in time is, obviously, achieved by complex conjugation of $U$.
%, provided the hamiltonian is real, ${\hat H}^* = {\hat H}$, which we will assume
 This brings us to the following choice of the the time-reversal operator ${\hat T}$ which is originally due to Wigner~\cite{wigner-tri}:
\begin{enumerate}
 \item ${\hat T}$ is an anti-unitary operator, i.e. ${\hat T} = {\hat K}{\hat W}$, where ${\hat W}$ is a unitary operator that assures the spin flip in Eq.~(\ref{eq:probability_reversal}) and ${\hat K}$ performs complex conjugation. Here we will not treat the spin explicitly and, therefore, use ${\hat W} \to 1$. As a result, Eq.~(\ref{eq:psi-psiprime}) turns into 
   \begin{equation}
     |\psi'(t')\rangle = \hat{T}|\psi(t)\rangle = |\psi(-t)\rangle^*,
    \label{eq:psiprime_final}   
    \end{equation}
 \item An operator ${\hat A}'$  acting on the time-reversed solution is obtained from the original operator ${\hat A}$ via 
  \begin{equation}
   {\hat A}' = {\hat T} {\hat A} {\hat T}^{-1}
  \label{eq:operator-transform}
  \end{equation}

 \item ${\hat T}$ is anti-linear, i.e. 
  \begin{align}
  {\hat T} \left\{|\psi_1\rangle + \i |\psi_2\rangle \right\} & = {\hat T} |\psi_1\rangle - \i {\hat T} |\psi_2\rangle,
  \label{eq:t-linearity1}
  \\   
 {\hat T} \left\{ {\hat A} + \i{\hat B} \right\}{\hat T}^{-1} &= {\hat T} {\hat A}{\hat T}^{-1} -\i {\hat T} {\hat B}{\hat T}^{-1},
  \label{eq:t-linearity2}
  \end{align}
for any two states, and any two operators. 
\end{enumerate}
As a test, we apply the operator $\hat{T}$ to both sides of Eq.~(\ref{eq:tdTSE}):
\begin{align}
 \hat{T}\,\i\hbar\,\partial_t|\psi\rangle &= \hat{T}\,\hat{H}|\psi\rangle\nonumber\\
 \Longleftrightarrow \; \underbrace{-\i\hbar\,\partial_t}_{\i\hbar\partial_{(-t)}}\hat{T}|\psi\rangle &= \hat{T}\hat{H}\hat{T}^{-1}\hat{T}|\psi\rangle\;,
\end{align}
which means that, indeed, $|\psi'\rangle = \hat{T}|\psi\rangle$ solves the time reversed Schr\"odinger equation 
\begin{equation}
 \i\hbar\, \partial_{(-t)}|\psi'\rangle = \hat{H}|\psi'\rangle
\end{equation}
if and only if
\begin{equation}
 \hat{H}=\hat{T}\,\hat{H}\,\hat{T}^{-1}.
\label{eq:h-condition}
\end{equation}
This is equivalent to $[\hat{T},\hat{H}] = 0$, and we recover a result found in many text books. However, we will see in Sec.~\ref{s:tri-bbgky} that condition (\ref{eq:h-condition}) is, in fact, not sufficient. 

Next, we find the time-reversed of the coordinate and momentum operators, using the coordinate representation, 
\begin{align}
 {\hat r}' = {\hat T}\, {\hat r}\,{\hat T}^{-1} = {\hat r}\,{\hat T} {\hat T}^{-1} = {\hat r}, 
\end{align}
since ${\hat r}$ is real, and 
\begin{align}
 {\hat p}' = {\hat T}\, {\hat p}\,{\hat T}^{-1} = - {\hat p}, 
 \label{eq:p-reversal}
\end{align}
since ${\hat p} = \frac{\hbar}{\i}\nabla$ is purely imaginary. This is again consistent with the time reversal properties of classical mechanics. Further, Eq.~(\ref{eq:p-reversal}) also shows that relation (\ref{eq:h-condition}) excludes certain classes of hamiltonians such as those containing odd powers of the momentum.

% This result is valid for an arbitrary interacting many-particle system. 

%-------------------------
\section{Time Reversal Invariance of the BBGKY-Hierarchy}\label{s:tri-bbgky}
%-------
%\subsection[Heisenberg Equation]{Time Reversal Invariance of the von Neumann Equation for the Density Operator ${\hat %{\rho}}_N$}
%\label{Heisenberg}
%
The $N$-particle density operator, ${\hat \rho}$, defined by Eq.~(\ref{eq:def-rho}), extends the concept of the time-dependent Schrödinger equation to a thermodynamic ensemble, while containing the dynamics of a pure state $|\psi^{(l)}\rangle$ as a special case, when $W_k = \delta_{k,l}$.
  
Let us now analyze the time reversal symmetry of the von Neumann equation (\ref{eq:von_neumann}), by applying
% is equivalent to the Schr\"odinger equation and should, therefore, possess the same reversibility properties. 
 the ${\hat T}$-operator introduced above from the left and its inverse from the right:
\begin{align}
% \i\hbar \,\partial_t\hat{\rho} &= \left[ \hat{\rho}, \hat{H}\right] \label{eq:HE}\\
% \Longleftrightarrow \; 
\hat{T}\,\i\hbar\,\partial_t\hat{\rho}\, \hat{T}^{-1} &= \hat{T}\left( \hat{\rho}\hat{H} - \hat{H}\hat{\rho}\right)\hat{T}^{-1} \nonumber\\
% \Longleftrightarrow \; 
-\i\hbar\,\partial_t\hat{T}\hat{\rho} \hat{T}^{-1} &= \hat{T}\hat{\rho}\hat{T}^{-1}\,\hat{T}\hat{H}\hat{T}^{-1} - \hat{T}\hat{H}\hat{T}^{-1}\,\hat{T}\hat{\rho}\hat{T}^{-1}\;,
\nonumber
\end{align}
which is equivalent to the time-reversed equation
\begin{equation}
%\i\hbar\,\partial_{-t}\hat{T}\hat{\rho} \hat{T}^{-1} = \left[\hat{T}\hat{A}_{\H}\hat{T}^{-1}, \hat{H}\right]
\i\hbar\,\partial_{-t}\,\hat{\rho}' = \left[\hat{\rho}', \hat{H}\right],
\end{equation}
if and only if 
again condition (\ref{eq:h-condition}) is fulfilled, as in the case of the Schrödinger equation. 
Here we introduced the solution of the time reversed von Neumann equation,
\begin{align}
 {\hat \rho}'(-t) &= \hat{T}\,\hat{\rho}(t)\,\hat{T}^{-1}
\\ \nonumber
&= \sum_k W_k \, \hat{T}|\psi^{(k)}(t)\rangle \langle \psi^{(k)}(t)|  \hat{T}^{-1}
\\ \nonumber
&= \sum_k W_k \, |\psi^{(k)'}(-t)\rangle \langle \psi^{(k)'}(-t)|, 
\end{align}
which is consistent with the definition of the density operator (\ref{eq:def-rho}) in terms of the  solutions of the time-reversed Schrödinger equation.
\begin{figure}
 \includegraphics[width=.5\linewidth]{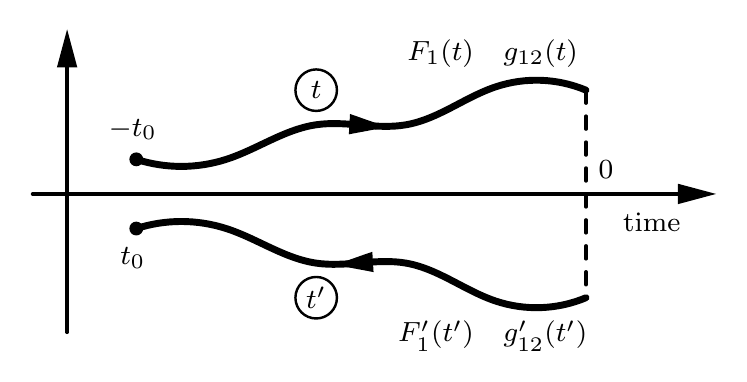}
 \caption{Illustration of the forward and backward solutions of the quantum BBGKY-hierarchy. Upper trajectory: forward solution $\{F_1(t), g_{12}(t), \dots\}$ on the interval $-t_0 \le t \le 0$. Lower trajectory: backward solution $\{F'_1(t'), g'_{12}(t'), \dots\}$ on the same interval with $0 \le t' \le t_0$, Time reversal occurs at $t=0$, cf. Fig.~\ref{fig:tri-tdse}.
 }
 \label{fig:tri-bbgky}
\end{figure}
Let us now return to the BBGKY-hierarchy (\ref{eq:bbgky_finite}). Its time-reversibility follows immediately from the reversibility of the von Neumann equation (\ref{eq:von_neumann}) that was demonstrated above. Nevertheless, it is instructive to verify the time-reversibility explicitly as this will be useful for the analysis of approximations to the hierarchy in Sec.~\ref{s:tri-approx}. Applying the operators $\hat{T}$ and $\hat{T}^{-1}$ from the left and right, respectively, we obtain
\begin{align}
%\label{eq:bbgky_finite}
& \i\hbar\frac{\partial}{\partial (-t)} {\hat F}'_{1\dots s}-[{\hat H}'_{1\dots s}(-t),{\hat F}'_{1\dots s}(-t)]
\nonumber\\
& =
 \mbox{Tr}_{s+1}\sum_{i=1}^{s}[{\hat V}'_{i,s+1},{\hat F}'_{1\dots s+1}(-t)],
\end{align}
where we used the fact that the definition (\ref{eq:fs-def}) is a real linear operation
\begin{align}
  \hat{T} \,\hat{F}_{1\dots s}(t)\, \hat{T}^{-1} &= C_s^N \mbox{Tr}_{s+1\dots N} \hat{T} \hat{\rho}(t)\,\hat{T}^{-1} = 
\nonumber \\
 &= C_s^N \mbox{Tr}_{s+1\dots N} \hat{\rho}'(-t) = \hat{F}'_{1\dots s}(-t),
\label{eq:fs-tri}
\end{align}
such that $\hat{F}'_{1\dots s}(-t)$ is, indeed, the solution of the time reversed hierachy equation, if the following conditions hold
\begin{align}
 {\hat H}'_{1\dots s}(-t) &\equiv  \hat{T} \,\hat{H}_{1\dots s}(t)\, \hat{T}^{-1} =  {\hat H}_{1\dots s}(t),
\label{eq:hs-tri}
\\
{\hat V}'_{ij} &\equiv  \hat{T} \,\hat{V}_{ij}\, \hat{T}^{-1} =  {\hat V}_{ij}, 
\label{eq:vij-tri}
\end{align}
for all $i \ne j \in [1, N]$ and all $s=1\dots N-1$, simultaneously. While for typical distance-dependent real potentials, Eq.~(\ref{eq:vij-tri}) is always fulfilled, Eq.~(\ref{eq:hs-tri}) puts clear restrictions on the contributions to the system hamiltonian.

 Based on these results we conclude that time-reversal invariance of the exact BBGKY-hierarchy requires not only the time reversal symmetry of the full $N$-particle hamiltonian (\ref{hn_def}), as in the case of the Schrödinger equation, cf. condition (\ref{eq:h-condition}), but each of the contributions to the hamiltonian have to obey this symmetry separately. This is, of course, a much stronger condition than (\ref{eq:h-condition}).

%----------------------
\section{Time Reversal Invariance of Approximations to the hierarchy}\label{s:tri-approx}
\noindent
Since the solution of the BBGKY-hierarchy is usually possible only with suitable approximations, the important question arises which approximations retain the time reversal properties of the exact system. In the following we demonstrate that a very broad class of approximations retains time reversal invariance.
Thereby we will restrict ourselves to real-valued Hamiltonians,  $\hat{H}^*=\hat{H}$.

We start by rewriting the first two equations of the BBGKY-hierarchy in terms of the correlation operators, Eqs.~(\ref{eq:f1-eq}, \ref{eq:g12-eq}) in a different form,
\begin{align}
\i\hbar\frac{\partial}{\partial t}{\hat F}_1 &= {\hat J}_1 = {\hat J}^{\rm app}_1 + {\hat O}_1,
\label{eq:f1-approx}
\\
\i\hbar\frac{\partial}{\partial t}{\hat g}_{12} &= {\hat J}_{12} = {\hat J}^{\rm app}_{12} + {\hat O}_{12},
\label{eq:g12-approx}
\end{align}
where ${\hat J}_1$ and ${\hat J}_{12}$ comprise all remaining terms in Eqs.~(\ref{eq:f1-eq}, \ref{eq:g12-eq}). 
A decoupling approximation to the hierarchy can then be defined by specifying approximate expressions, ${\hat J}^{\rm app}_1$ and ${\hat J}^{\rm app}_{12}$, where the remainders, ${\hat O}_1$ and ${\hat O}_{12}$, are being omitted. The same procedure can be applied to decoupling approximations on the level of the third or higher order hierarchy equations.
To answer the question whether a given decoupling approximation, ${\hat J}^{\rm app} = \{ {\hat J}^{\rm app}_1, {\hat J}^{\rm app}_{12}, \dots \}$, is time reversible we either have to analyze the resulting equations directly or, alternatively, investigate the time reversal properties of the omitted operators, ${\hat O}=\{ {\hat O}_1, {\hat O}_{12}, \dots \}$, since the exact equations are known to be time reversal invariant. Below it will be advantageous to use the latter approach.

In the following we answer this question for the approximations that were introduced in Sec.~\ref{s:bbgky}, starting by specifying the corresponding operators ${\hat O}$.
\begin{enumerate}
 \item The {\it mean field approximation} is given by the choice ${\hat O}\equiv {\hat O}^{\rm HF}_1 = \mbox{Tr}_2 [{\hat V}_{12}, {\hat g}_{12}]$.
 \item The {\it second order Born approximation} is given by ${\hat O}\equiv {\hat O}_{12}^{\rm 2B} = [{\hat V}_{12}, {\hat g}_{12}] + \\
\mbox{Tr}_{3}\Big\{ [{\hat V}_{13},{\hat F}_1 {\hat g}_{23}] + [{\hat V}_{23},{\hat F}_2 {\hat g}_{13}] +
[{\hat V}_{13}+{\hat V}_{23},{\hat g}_{123}]\Big\}$.
  \item The {\it T-matrix or ladder approximation} is given by ${\hat O}\equiv {\hat O}^{\rm T}_{12} = \\ \mbox{Tr}_{3}\Big\{ [{\hat V}_{13},{\hat F}_1 {\hat g}_{23}] + [{\hat V}_{23},{\hat F}_2 {\hat g}_{13}] +
[{\hat V}_{13}+{\hat V}_{23},{\hat g}_{123}]\Big\}$.
  \item The {\it polarization approximation} is given by \\${\hat O}\equiv {\hat O}_{12}^{\rm POL} = [{\hat V}_{12}, {\hat g}_{12}] + \mbox{Tr}_{3} [{\hat V}_{13}+{\hat V}_{23},{\hat g}_{123}]$.
  \item The {\it screened ladder approximation} is given by\\ ${\hat O}\equiv {\hat O}_{12}^{\rm SCT} = \mbox{Tr}_{3} [{\hat V}_{13}+{\hat V}_{23},{\hat g}_{123}]$.
\end{enumerate}
Aside from their different physical character, all these approximations have a common mathematical structure. They are given by a functional relation of the form 
\begin{align}
\label{eq:def-o}
 {\hat O}(t) &= R[{\hat V}_{ij}, {\hat F}_k(t), {\hat g}_{lm}(t), {\hat g}_{nop}(t), \dots], \quad R \in \cal{R}, 
\end{align}
where the indices $i, j, k, l, m, n, o, p \in 1\dots N$, and $R$ is a real function. The properties of expression (\ref{eq:def-o}) under time reversal are easily obtained. First, due to its real character, obviously, the functional form of $R$ does not change, i.e. ${\hat T} R {\hat T}^{-1} = R$. Second, the properties of the arguments of $R$ are known: As we have discussed above, standard pair potentials are always time reversal invariant, $\hat{T} \,\hat{V}_{ij}\, \hat{T}^{-1} =  {\hat V}_{ij}$. Next, time reversal invariance of the single-particle density operator was demonstrated in Eq.~(\ref{eq:fs-tri}). Finally, property (\ref{eq:fs-tri}), together with the cluster expansion (\ref{eq:g12-def}, \ref{eq:g123-def}, \dots), which is a real functional relation, we easily conclude (iteratively) that all correlation operators are time reversal invariant, 
\begin{align}
 \hat{T} \,\hat{g}_{1\dots s}(t)\, \hat{T}^{-1} =  {\hat g}_{1\dots s}(-t), \quad s=1\dots N-1.
\end{align}
Summarizing these results we conclude that the operator (\ref{eq:def-o}) is time reversal invariant, 
\begin{align}
 \hat{T} \,\hat{O}(t)\, \hat{T}^{-1} =  {\hat O}(-t).
\end{align}
This means that each of the approximations that were listed above (and the corresponding non-Markovian quantum kinetic equations)---time-dependent Hartree-Fock (nonlinear quantum Vlasov equation), second order Born approximation (quantum Landau equation), T-matrix (quantum Boltzmann equation), polarization approximation (quantum Len\'ard-Balescu equation) and the screened ladder approximation---are time reversal invariant. 
We underline that the condition (\ref{eq:def-o}) is much more general than those approximations, including a broad range of decoupling schemes of the hierarchy that were proposed in the literature.

\section{Breaking the time-reversal symmetry: example of the Born approximation}\label{s:born}
The emergence of time irreversibility, starting from reversible quantum dynamics has been discussed in great detail since the appearance of Boltzmann's kinetic equation~\cite{boltzmann72}. Using our formalism, we can trace this emergence particularly clearly for the case of the quantum Landau equation that corresponds to the following first two hierarchy equations: 
\begin{align}
& \i \hbar\frac{\partial}{\partial t} {\hat F}_{1} - [{\hat {\bar H}_1},{\hat F}_1]
=
\mbox{Tr}_{2}[{\hat V}_{12},{\hat g}_{12}],
\label{eq:f1-eq2b}
\\
& \i\hbar \frac{\partial}{\partial t}
{\hat g}_{12} - [{\hat {\bar H}}_{12}, {\hat g}_{12}]
= [{\hat V}_{12},{\hat F}_{1}{\hat F}_{2}]^\pm = {\hat J}^{2B}_{12}(t),
\label{eq:g12-eq2b}
\\
&{\hat F}_1(-t_0) = {\hat F}^0_1, \quad {\hat g}_{12}(-t_0) = {\hat g}^0_{12}, \quad t\in [-t_0, 0],
\label{eq:2bic}
\end{align}
where we added the initial conditions for both operators. These coupled time-local equations can be solved directly numerically. The alternative route that leads to a quantum kintic equation consists in,  first,  formally solving the equation for ${\hat g}_{12}$ analytically and then inserting the result into the r.h.s. of Eq.~(\ref{eq:f1-eq2b}). This is the approach we will use here. The solution of the initial value problem (\ref{eq:f1-eq2b}--\ref{eq:2bic}) is easily found~\cite{bonitz_qkt} and consists of an initial value term (solution of the homogeneous equation) and a collision term
\begin{align}\label{eq:g12-sol}
 {\hat g}_{12}(t) &= {\hat g}^{\rm IC}_{12}(t) + {\hat g}^{\rm coll}_{12}(t),
\\
{\hat g}^{\rm IC}_{12}(t) &= {\hat U}_{12}^0(t,-t_0) \, {\hat g}^0_{12} \,{\hat U}_{12}^{0\dagger}(t,-t_0),
\\
{\hat g}^{\rm coll}_{12}(t) &= \frac{1}{\i\hbar} \int\limits_{-t_0}^t \d{\bar t} \,
{\hat U}_{12}^0(t,{\bar t}) \, {\hat J}^0_{12}({\bar t}) \,{\hat U}_{12}^{0\dagger}(t,{\bar t}),
\end{align}
where the two-particle propagator factorizes into single-particle Hartree-Fock propagators, ${\hat U}_{12}^0(t,t') = {\hat U}_{1}(t,t'){\hat U}_{2}(t,t')$, with
\begin{align}
 \left\{\i\hbar \frac{\partial}{\partial t} - {\hat {\bar H}}_1(t) \right\}{\hat U}_{1}(t,t') = 0, \quad {\hat U}_{1}(t,t)=1,
\label{eq:u2born-equation}
\end{align}
the solution for which is analogous to that for the Schrödinger equation, cf.~Eq.~(\ref{eq:un-solution}). 
The quantum kinetic equation that is associcated with the solution (\ref{eq:g12-sol}) contains two collision integrals: the first, involving ${\hat g}^{\rm IC}_{12}(t)$, is due to correlations existing in the system at the initial time moment whereas the second is due to correlations being formed as a result of two-particle collisions while being absent at the initial moment. The characteristic feature of the latter collision integral is its non-Markovian character (i.e. the presence of the time integral) which is in striking contrast to the traditional Boltzmann equation that involves only distribution functions taken at the current time $t$. 

To analyze the transition from the former to the latter and, thereby, from time reversibility to irreversibility we switch from the operator form of the solution (\ref{eq:g12-sol}) to an instantanteous Hartree-Fock basis $\{|n \rangle \}$ given by ${\hat {\bar H}}_1|n\rangle = E_n |n\rangle$. Then the first hierarchy equation (\ref{eq:f1-eq2b}) becomes
\begin{align}\label{eq:qkin-eq}
 \i\hbar\frac{\partial}{\partial t}F_{n_1, n_1'} &- \left( E_{n_1}-E_{n'_1} \right)F_{n_1, n_1'} =
\\
&= \sum_{n_2}\sum_{{\bar n}_1{\bar n}_2}\left\{ V_{n,{\bar n}}g_{{\bar n},n'} - g_{n,{\bar n}}V_{{\bar n},n'}\right\}\big|_{n_2'=n_2},
\nonumber
\end{align}
where we introduced the short notations $n \equiv (n_1, n_2)$, $n' \equiv (n'_1, n'_2)$ and  ${\bar n} \equiv ({\bar n}_1, {\bar n}_2)$. This is a generalized quantum kinetic equation that describes the probability of transitions between different single-particle states (dynamics of $F_{n_1, n_1'}$ with $n_1 \ne n_1'$), as well as the dynamics of the occupations of state $n_1$ (given by $F_{n_1}\equiv F_{n_1, n_1}$). Here we focus on the latter as it is directly related to the evolution towards an equilibrium state. Further, emergence of irreversibility in the dynamics of $F_n$ is sufficient for the transition of the whole system of coupled equations from reversible to irreversible.

The corresponding dynamics of the diagonal matrix elements are given by 
\begin{align}
 \i\hbar\frac{\partial}{\partial t}F_{n_1}(t) = 2\i
\sum_{n_2}\sum_{{\bar n}_1{\bar n}_2} V_{n,{\bar n}}\, {\rm Im}\,g_{{\bar n},n}(t),
\label{eq:diagonal-f1eq}
\end{align}
where we used $g_{n,n'}=g^*_{n',n}$ and $V_{n,n'}=V_{n',n}$. 
%{\bf das gilt in der Impulsdarstellung, ist das korrekt in einer beliebigen Darstellung??} 
To compute ${\rm Im}\,g_{{\bar n},n}(t)$, we first write down the solution of Eq.~(\ref{eq:u2born-equation}) which is given by a diagonal matrix
\begin{align}
 \langle n_1| {\hat U}(t,t')|n_1'\rangle &= U_{n_1}(t-t')\, \delta_{n_1,n_1'},
\nonumber\\
U_{n_1}(\tau) &= e^{-\frac{\i}{\hbar}E_{n_1} \tau},
\end{align}
 and the matrix of the pair correlation operator (\ref{eq:g12-sol}) becomes
\begin{align}\label{eq:g12-sol_basis}
{\rm Im}\, g_{n, n'}(t) &= {\rm Im}\,g^{\rm IC}_{n,n'}(t) + {\rm Im}\,g^{\rm coll}_{n,n'}(t),
\\
{\rm Im}\,g^{\rm IC}_{n,n'}(t) &= {\rm Im}\,\left\{ e^{-\i\omega_{n,n'}[t-(-t_0)]}\, g^0_{n,n'}\right\} \,,
\label{eq:gic-solution-bases}\\
{\rm Im}\,g^{\rm coll}_{n,n'}(t) &= -\frac{1}{\hbar} \int\limits_{-t_0}^t \d{\bar t} \,
 \cos{[\omega_{n,n'}(t-{\bar t})]} \, J^{\rm 2B}_{n,n'}({\bar t}) \,,
\label{eq:gcol-solution-bases}
\end{align}
where we defined $\hbar \omega_{n,n'} \equiv E_{n_1}+E_{n_2}-E_{n'_1}-E_{n_2}'$ and used $J^{\rm 2B*}_{n,n'}=J^{\rm 2B}_{n,n'}$.

Let us now investigate the time reversal symmetry of the kinetic equation (\ref{eq:qkin-eq}), i.e., we apply the time reversal operators ${\hat T}$ and ${\hat T}^{-1}$, from the left and right, respectively, as before,
\begin{align}
%\label{eq:qkin-eq}
\nonumber
 \quad & \i\hbar\frac{\partial}{\partial (-t)}F'_{n_1, n_1'}(t) - \left( E_{n_1}-E_{n'_1} \right)F'_{n_1, n_1'}(t) =
\\
&= \sum_{n_2}\sum_{{\bar n}_1{\bar n}_2}\left\{ V_{n,{\bar n}}g'_{{\bar n},n'}(t) - g'_{n,{\bar n}}(t)V_{{\bar n},n'}\right\}\big|_{n_2'=n_2},
\nonumber
\end{align}
where $F'$ is the solution of the time-reversed equation. Time reversal symmetry again requires fulfillment of
$F'_{n_1, n_1'}(t)\equiv {\hat T} F_{n_1, n_1'}(t) {\hat T}^{-1} = F_{n_1, n_1'}(-t)$ and is observed only when the time reversed solution of the second equation obeys 
\begin{align}
{\rm Im}\, g'_{n, n'}(t)\equiv {\hat T} {\rm Im}\,g_{n_1, n_1'}(t) {\hat T}^{-1} = {\rm Im}\,g_{n, n'}(-t).
\label{eq:g2b-reversibility}
\end{align}
This is easily verified by writing down the solution $g'(t)$ noticing that application of the operators ${\hat T}$ and ${\hat T}^{-1}$, from the left and right to the second hierachy equation again changes the sign of the time derivative which is equivalent to replacing $\omega_{n,n'} \to - \omega_{n,n'}$ and $J^{\rm 2B}_{n,n'} \to J^{\rm 2B}_{n,n'}$, and the solution (\ref{eq:gic-solution-bases}, \ref{eq:gcol-solution-bases}) changes into
\begin{align}
{\rm Im}\,g^{\rm IC\,'}_{n,n'}(t) &= {\rm Im}\,\left\{ e^{+ \i\omega_{n,n'}[t-(-t_0)]}\, g^0_{n,n'}\right\} \,,
\label{eq:gprimeic-solution-bases}\\
{\rm Im}\,g^{\rm coll\,'}_{n,n'}(t) &= -\frac{1}{\hbar} \int\limits_{-t_0}^t \d{\bar t} \,
 \cos{[-\omega_{n,n'}(t-{\bar t})]} \, [- J^{\rm 2B}_{n,n'}({\bar t})] \,,
\nonumber\\
= -\frac{1}{\hbar} & \int\limits_{-t}^{t_0} \d{\bar t} \,
 \cos{[\omega_{n,n'}(-t-{\bar t})]} \, J^{\rm 2B}_{n,n'}({- \bar t}) \,.
\label{eq:gprimecol-solution-bases}
\end{align}
It is obvious that the solutions $g$ and $g'$ fulfill (\ref{eq:g2b-reversibility}) which is seen by changing $(t, -t_0) \to (-t, t_0)$, in $g^{\rm IC\,'}$, and $(t, -t_0, {\bar t}) \to (-t, t_0, -{\bar t})$, in $g^{\rm coll\,'}$.
% (using $t_0 = 0$).

The mathematical transition to the conventional (quantum) Boltzmann collision integral that contains a delta function, $\delta(E_{n_1}+E_{n_2}-E_{n'_1}-E_{n'_2})$, of the single-particle energies before and after the collision involves three steps:
\begin{enumerate}
 \item {\it Decoupling of the time scales of the single-particle and two-particle dynamics}. The argument here is that, during a collision when the two-particle correlations are formed (during the correlation time $\tau_\mathrm{cor}$) the occupation of the single-particle states changes only weakly. Its relaxation towards an equilibrium distribution involves many collisions and, therefore, requires a relaxation time that is much larger, 
\begin{equation}
t_\mathrm{rel} \gg \tau_\mathrm{cor}.
\label{eq:trel-taucor} 
\end{equation}
This justifies to expand, $F_{n}({\bar t})$, and with it $J^{\rm 2B}_{n,n'}({\bar t})$, under the time integral in (\ref{eq:gcol-solution-bases}) around its value at the upper limit (the current time), 
\begin{align}\label{eq:retardation-exp}
 J^{\rm 2B}_{n,n'}({\bar t}) = J^{\rm 2B}_{n,n'}(t) + \sum_{k=1}\frac{({\bar t}-t)^k}{k!}\frac{\d^k}{\d t^k} J^{\rm 2B}_{n,n'}(t).
\end{align}
Truncating this retardation expansion~\cite{bonitz_qkt} at the first term ($0$-th order retardation approximation) leads to the following result for the pair correlations,
\begin{align}
 {\rm Im}\, g^{\rm coll\,(0)}_{n,n'}(t) &= -\frac{J^{\rm 2B}_{n,n'}(t)}{\hbar} \;
\frac{ \sin{[\omega_{n,n'}(t-(-t_0))]} }{\omega_{n,n'}}
\nonumber\\
&= {\rm Im}\, g^{\rm coll\,(0)}_{n,n'}(t, [F(t)]).
\label{eq:gcoll-0}
\end{align}
This expression is, of course, a drastic distortion of the original result and its accuracy depends on the fulfillment of condition (\ref{eq:trel-taucor}). In fact, it is well known that, for weakly coupled systems, the two times are related by $\frac{\tau_\mathrm{cor}}{t_\mathrm{rel}}\sim \Gamma \ll 1$, where $\Gamma$ is the relevant coupling parameter. 
In the second line of (\ref{eq:gcoll-0}) we noted explicitly that the pair correlation functions have a two-fold time dependence: an explicite one (via the sine function, which is fast, for increasing time, in particular for high frequencies) and a slow one--via the evolution of $F(t)$.

Note that this is still a proper (although distorted) solution of the initial value problem. It is also consistent with an (arbitrary) initial condition $g^0_{n,n'}(-t_0)$, because the collision term exactly vanishes for $t\to -t_0$.
Interestingly, despite the approximate character of $g^{\rm coll\,(0)}_{n,n'}(t)$, it is easily seen [by performing the retardation expansion in (\ref{eq:gprimecol-solution-bases})] that it still satisfies the time reversal invariance condition (\ref{eq:g2b-reversibility}).
\item {\it Markov limit}. The limit of an infinitely remote initial state, $-t_0 \to -\infty$ is usually motivated by the assumption that two particles enter a scattering process in and uncorrelated manner.
The result for the Markovian pair correlations is then
\begin{align}
 {\rm Im}\, g^{\rm coll\,(M)}_{n,n'}(t) & \equiv
\nonumber\\
-\frac{J^{\rm 2B}_{n,n'}(t)}{\hbar} &\;\lim_{-t_0 \to -\infty}
\frac{ \sin{[\omega_{n,n'}(t-(-t_0))]} }{\omega_{n,n'}}
\nonumber\\
& = -\frac{J^{\rm 2B}_{n,n'}(t)}{\hbar} \;
\delta(\omega_{n,n'}).
\label{eq:gcoll-M}
\end{align}
Note that it is assumed that the single-particle operators (i.e. the slow time-dependence of $g^{\rm coll\,(0)}$) are not affected by the limit which means that first the limit $\Gamma \to 0$ has been taken.
\item {\it Weakening of initial correlations}. Motivated by the argument that the state of the system cannot remember (and, hence, be influenced by) its infinitely remote history and, in particular, its correlations, the Markov limit is accompanied by the suppression of initial correlations, 
\begin{align}
 \lim_{-t_0 \to -\infty}  g^{0}_{n,n'}(-t_0) \longrightarrow 0.
\label{eq:ic-weakening}
\end{align}
This is consistent with the Markov limit because, after the procedure leading to (\ref{eq:ic-weakening}),  $g_{n, n'}[F(t)]$ does not obey an initial value problem anymore that starts from an arbitrary initial state, but only adiabatically follows the dynamics of $F(t)$, according to the prescription (\ref{eq:gcoll-M}). 
This concept is due to Bogolyubov~\cite{bogolyubov} (``functional hypothesis''; ``weakening of initial correlations'') and has been generalized to situations where there exists a subclass of long-living correlations (such as those related to bound states or long range order; {\it partial weakening of initial correlations}) by Kremp {\it et al.}~\cite{kremp81}.

With the result (\ref{eq:ic-weakening}) the collision integral due to initial correlations (the term $g^{\rm IC}$) vanishes and only the collision integral involving ${\rm Im}\, g^{\rm coll\,(M)}_{n,n'}$, Eq.~(\ref{eq:gcoll-M}) remains which has the convential Boltzmann-type form. 
\end{enumerate}
To summarize, time reversal symmetry is lost at step 2. While the result of step 1, ${\rm Im}\, g^{\rm coll\,(0)}_{n,n'}(t)$, is time-reversal invariant {\it for any finite value} $-t_0$, no matter how far back in the past, this property vanishes with the limit $-t_0 \to -\infty$. With this limit the unitary operator structure that is still present in the sine function is lost together with the explicit time dependence of the pair correlations (this is particularly clear when the single-particle operators $F$ are exactly stationary.)

%-------------------
\section{Summary and Discussion}\label{s:summary}
In this paper we analyzed the question of time reversibility of generalized quantum kinetic equations that are derived within the reduced density operator formalism. The governing equations of the density operator theory are 
given by the quantum BBGKY-hierarchy. Here we demonstrated that the exact BBGKY-hierarchy and the associated quantum-kinetic equations are time reversible.
This behavior is in striking contrast to conventional Boltzmann-type kinetic equations that are known to be irreversible and describe the relaxation of a many-body system to an equilibrium state which is accompanied by an increase of its entropy (H-theorem).
This is traditionally achieved by means of ad hoc assumptions such as about ``molecular chaos''~\cite{mol-chaos-maxwell}, via Boltzmann's ``Stoßzahlansatz''~\cite{boltzmann72} or by similar procedures. 

Although the derivation of generalized non-Markovian quantum kinetic equations goes back almost seven decades, in many communities the existence a systematic kinetic theory beyond the Boltzmann equation is poorly known which warrants a detailed reconsideration of some mathematical aspects on the way from a reversible to an irreversible kinetic theory. Here we have presented a simple procedure that allows one to directly verify the time reversal property of the exact BBGKY-hierachy and of important closure relations, as well as the transition to the conventional Boltzmann equation. Our approach is based on the use of Wigner's anti-unitary time-reversal operator ${\hat T}$~\cite{wigner-tri} that translates the solution of the Schrödinger equation into the time-reversed equation and is a mathematically well controlled procedure that replaces the traditional heuristic arguments mentioned above.

Let us summarize our main results: 
\begin{enumerate}
 \item Our proof of time reversal invariance of the exact quantum BBGKY hierarchy revealed a much stronger condition, Eqs.~(\ref{eq:hs-tri}, \ref{eq:vij-tri}), than  the commonly used condition for time reversibility of the $N$-particle Schrödinger equation, i.e. Eq.~(\ref{eq:h-condition}). We have shown that not only the total hamiltonian has to obey ${\hat T}{\hat H}{\hat T}^{-1}={\hat H}$, but each of its single-particle, two-particle and higher contributions, separately. This might seem surprising since Eq.~(\ref{eq:h-condition}) is known to be necessary and sufficient for the Schrödinger and von Neumann equations. However,  the $N$-particle dynamics have always to be consistent with the quantum dynamics of sub-complexes (of $N-1\dots 1$ particles) which follow directly from partial integration of the $N$-particle equations. It is clearly impossibile that the $N$-particles dynamics are reversible whereas the $N$-$s$-particle dynamics are not.
 \item We presented a very general condition for time-reversal invariance of approximate solutions to the BBGKY-hierarchy, Eq.~(\ref{eq:def-o}), and showed that it applies to many of the commonly used many-body approximations. Moreover, this condition goes far beyond those approximations, including a broad range of additional decoupling schemes of the hierarchy. This is not limited to approximations that are motivated by physical considerations and violate conservation laws. For example, the choice of the omitted term ${\hat O}={\hat O}_{12} \ne {\hat O}_{21}$ would violate conservation of total energy, cf. Ref.~\onlinecite{bonitz_qkt}, while still being time reversal invariant.
 \item Our results allow us to analyze the interesting question posed in Ref.~\onlinecite{scharnke_17} of how total energy conservation and time reversibility are related. While in most cases of practical relevance both phenomena are fulfilled (or violated) simultaneously, their areas of validity are not equivalent. As shown above, there exist time-reversible models that violate energy conservation. On the other hand, there exist model hamiltonians (e.g. those that contain odd powers of the momentum) that conserve energy but violate condition (\ref{eq:h-condition}) and, therefore, time-reversal symmetry. 
 \item Our analysis of the transition to the conventional Boltzmann equation involved three successive approximations. The first one--the decoupling of the relaxation time scales of single-particle $t_\mathrm{rel}$ and two-particle dynamics ($\tau_\mathrm{cor}$) by means of a retardation expansion--allowed us to perform the memory integral and obtain a time-local result for the pair correlations, Eq.~(\ref{eq:gcoll-0}). This result (``completed collision approximation'' or ``energy broadening approximation'') not only conserves total energy~\cite{bonitz_qkt}, but here we also showed that it preserves time reversal symmetry. The same analysis also applies to higher order approximations in the retardation expansion~(\ref{eq:retardation-exp}).
 \item We have demonstrated that time reversibility is lost only at the second step--the Markov limit, i.e. with the shift of the initial time to the infinitely remote past, $-t_0 \to -\infty$. This destroys the unitary character of the dynamics of the pair correlations and introduces a preferred ``arrow'' of time because there is no way the system can ever return into this state.
 \item Our analysis also shows that the commonly used argument that irreversibility is introduced into the theory via the assumption of ``molecular chaos``~\cite{mol-chaos-maxwell} or the ``Stoßzahlansatz''~\cite{boltzmann72}, has to be stated with some care. The requirement that the two-particle probabilites factorize and particles enter the collision uncorrelatedly---i.e., in our notation $F_{12}=F_1 F_2$ or $g_{12}\equiv 0$---is not sufficient. First, transition to irreversibility is also possible in a strongly correlated system where this factorization is not possible, e.g. Ref.~\onlinecite{bonitz_qkt}. Second, the example of the Born approximation that we discussed in Sec.~\ref{s:tri-approx} applied to a weakly coupled system. Choosing, as the initial condition an uncorrelated system, i.e. $g(-t_0)=g^0 = 0$, we would formally satisfy those assumptions. Nevertheless, the resulting dynamics, would still be given by Eq.~(\ref{eq:g12-sol}) without the initial correlation term, but it would be perfectly time reversible. The crucial point for the emergence of irreversibility is again that the factorization is introduced not at a finite initial time, but in the infinitely remote past.
\end{enumerate}

Having the generalized quantum kinetic equations that were discussed above at our disposal, one may ask whether it is necessary at all to force the transition to conventional irreversible Boltzmann-type kinetic equations, given the rather crude approximations involved. The argument for the latter has always been that macroscopic many-particle dynamics such as transport (diffusion, heat conduction, viscosity, fluid dynamics etc.) is dissipative, and the dynamics are expected to approach thermodynamic equilibrium--the state of maximum entropy. The answer is clearly ``No''. Experience in solving the generalized quantum kinetic equations (e.g. Ref.~\onlinecite{bonitz_qkt}), that are derived either from the BBGKY-hierarchy or from nonequlibrium Green functions for a sufficiently long time clearly reveals that these solutions exhibit an irreversible trend towards an asymptotic state that is consistent with thermodynamic equilibrium. However, this state is different from a Maxwellian, Fermi or Bose momentum distribution as a result of correlations. Certainly, the present reversible dynamics will return to the initial state, however the associated Poincar\'e recurrence time increases exponentially with particle number. This behavior is in complete agreement with simulation results for classical systems: solutions of the reversible equations of classical mechanics of a many-particle system by means of microcanonical molecular dynamics show perfect relaxation trends to (correlated) thermodynamic equilibrium. 

Therefore, the choice between the irreversible Boltzmann-type kinetic equations and reversible generalized kinetic equations is mainly governed by the substantially increased computational effort involved in the solution of the latter.
Here, in fact, proof of time-reversibility of the relevant approximations that was given in this paper, is of high practical value as it provides a sensitive test for the numerical accuracy and convergence, e.g. Ref.~\onlinecite{stan-comment}. Time reversibility is also of importance for ``echo''-type experiments (e.g. Loschmidt echo \cite{loschmidt}, spin echo, Rabi flop etc.) where time reversal is being forced by an external pulse. The analysis of the forward and backward dynamics gives important insights into the internal properties (e.g. dissipation channels) of a many-body system, and the present generalized quantum kinetic equations are well suited for such investigation. For a recent theoretical analysis, see Ref.~\onlinecite{kehrein16}

%\bibliography{time_reversal_theory}

%isolated systems, unitary dynamics

% relaxation in reversible dynamics, Poincare time

Acknowledgements: we acknowledge support from the Deutsche Forschungsgemeinschaft via grant BO1366-9. 
%and from DAAD (?). 

\end{document}